\documentclass[preprint,showpacs,showkeys,aps]{revtex4} %preprint
\begin{document}
\title{The Renormalization Group and Fractional Brownian Motion}
\author{David Hochberg}
\email{hochberg@laeff.esa.es} 
\author{Juan P\'erez--Mercader}
\email{mercader@laeff.esa.es}
\affiliation{Centro de Astrobiolog\'{\i}a, (CSIC-INTA), 28850 Torrej\'on
de Ardoz, Madrid, Spain.}
\date{\today}

\begin{abstract}
We find that in generic field theories the
combined effect of fluctuations and interactions leads to a
probability distribution function which describes fractional
Brownian Motion (fBM) and ``complex
behavior''. To show this we use the Renormalization Group as a tool
to improve perturbative calculations, and check that beyond the
classical regime of the field theory (i. e., when no fluctuations are
present) the non--linearities drive the probability distribution
function of the system away from classical Brownian Motion and into
a regime which to the lowest order is that of fBM. Our results can be
applied to systems away from equilibrium and to dynamical
critical phenomena. We illustrate our results with two selected examples:
a particle in a heat bath, and the KPZ equation. 
\end{abstract}



\maketitle

Complex behavior is ubiquitous \cite{science99}. From fluids to
ecosystems to chemistry, we are familiar with phenomena often
associated with non--gaussian Probability Distribution Functions
(PDFs); phenomena described by PDFs with ``long tails'' or ``stretched
exponential'' behaviors \cite{shlesingerphysicstoday}. The PDFs
associated with these phenomena have the property that, typically,
\cite{goldenfeldandkadanoff} ``improbable (very bad) events are much
more likely than with a Gaussian''. Many of these
phenomena are more or less loosely associated with, e. g.,
``complexity phase transitions'' \cite{kauffmann} and ``fractal
behavior'' \cite{mandelbrot}; furthermore, often, ``what is seen
depends on the size of the observer'' \cite{goldenfeldandkadanoff} and
is accompanied by the property that ``one law leads to many
behaviors'' \cite{goldenfeldandkadanoff}. They occur in systems that
are extended in space and in time, with many interacting components
and where both, ``the random and the regular'' \cite{mgmlectures}, are
at work.

Here we show how this ``complex behavior'' can be understood as a
natural consequence of space--time evolution, fluctuations and
interactions in a many body system observed at different scales and/or
with varying initial conditions on the parameters describing the
system. We do this for {\it a generic} many body system: a (quantum,
thermal or stochastic) field theory where space--time evolution and
interactions represent the ``regular'' and fluctuations represent the
``random''.

The logic we use is the following: as is well known \cite{weinberg},
in any field theory fluctuations and interactions (a) modify the
$n$-point correlation functions of the field at different space-time
points, and (b), lead to divergences in the original set of parameters
such as couplings and diffusion constants, masses, etc. defining the
system.  From (a) it follows that the {\em characteristic function} of
the field theory is modified from its originally normal and gaussian
character, and therefore the PDF itself is also modified; furthermore,
(b) leads to a scale-dependence of the parameters and correlation
functions which, asymptotically, are power laws \cite{wilsonandfisher}
with exponents calculable within the framework of the Renormalization
Group (RG) \cite{gml}, \cite{goldenfeld}.

Specifically, we address here the problem of how
interactions and fluctuations manifest themselves at $asymptotic$
$scales$ on the system's probability distribution and discover that
(i) the specific asymptotic properties of the PDF depend on the basins
of attraction of various RG fixed points and (ii) that,
asymptotically, the field theory displays the type of behavior we have
called ``complex'' above.

We begin by introducing the random field $\phi(\vec x,t_x)$.  We
coarse-grain (or filter) this field by means of a ``window-function''
$W_R$ which averages out the small scale features in $\phi$; here we
mean small in comparison to a window length scale $R$. Without loss of
generality, we take the window function to be translationally
invariant and define

\begin{equation}\label{window}
\phi_R(y) = \frac{1}{V}\int dx\, W_R(y - x) \, \phi(x),
\end{equation}

\noindent
where it is understood that $x =(\vec x, t_x)$ where $\vec x$ is a
$d$-dimensional vector, and $V = \Omega T$ is the spatial volume
$\Omega$ times the time interval $T$ over which the field is filtered.
Typical window functions are the ``top-hat'' window $W_R(x) = \Theta(R
- |x|)$ and the Gaussian window $W_R(x) = \exp(-|x|^2/{2R^2})$.  In
(\ref{window}) the coarse-grained field $\phi_R(y)$ is the result of
averaging $\phi$ over a space-time region or ``cell'' of linear extent
$R$, ``centered'' at space-time point $y$, over the time interval $T$.
The small scale features of the field are blurred within the
space-time region selected by the window: the resolution is degraded.
We also note that (\ref{window}) is a linear transformation relating
the microscopic ($\phi$) to the coarse-grained field ($\phi_R$) that
need not be invertible.

We can calculate \cite{Itzykson,Rivers} the probability that the
coarse-grained field $\phi_R(y)$ takes the $value$ $\varphi$ within
the window:

\begin{equation}\label{pdf}
p(\varphi; R,y)= \langle \delta( \phi_R(y) - \varphi) \rangle_P \equiv 
\int [{\cal D}\phi]\, \delta( \phi_R(y) - \varphi)\, P[\phi]\,\, ,
\label{1}
\end{equation}

\noindent
where $P[\phi]$ is the PDF for the microscopic field configurations
and the integral is a path integral over all field
configurations. 
We point out that $\int d\varphi \, p(\varphi) = \int [{\cal D}\phi]\,
P[\phi] = 1$, so that $p$ is normalized to unity if and only if 
the PDF is. We see moreover that $p$ is a probability density and scales
dimensionally as the inverse field $\varphi^{-1}$. 
We will make use of this fact below. 
The above probability $p$ can be easily related to
$Z[J]$, the characteristic functional with source $J$ (more commonly
known as the generating functional) for the $n$--point functions. In
fact, by using the Fourier integral representation of the
delta distribution in Eq. (\ref{1}) we have

$$
p(\varphi; R, y)= \int d \xi \int [{\cal D}\phi] \, e^{i \xi [\phi_R
  (y) - \varphi]}
P[\phi]
$$
$$
= \int d \xi \, e^{-i \xi \varphi}\int [{\cal D}\phi] \, 
e^{i \frac{\xi}{V} \int dx 
  W_R (x - y) \phi
  (x)} P[\phi]
$$
\begin{equation}
=\int_{-\infty}^{\infty} d \xi \, e^{-i \xi \varphi} 
\, Z[J_R(x-y)= i\frac{\xi}{V} W_R(x-y) ]\,\, ,
\label{2}
\end{equation}
 
\noindent
where $Z$ in the last line is the $function$ of $\xi$ that results
from evaluating the generating functional for a window-source
function.  For a $generic$ gaussian probability functional
\cite{weinberg}, \cite{spde} and \cite{zinnjustin}

\begin{equation}
P[\phi] = {\cal N} \exp \left\{ - \frac{1}{2} \int \, dx\, \int  dz\,
\phi(x) G^{-1}(x,z)
\phi(z)
\right\}\,\,\, ,
\label{3}
\end{equation}

\noindent
the path integral in Eq.(\ref{2}) is gaussian (${\cal N}$ is
a normalization factor) and one immediately obtains

\begin{equation}\label{zeroth}
p(\varphi; R, y)=\sqrt{\frac{V^2}{2\pi[W_R\cdot G \cdot W_R] }}
\cdot \exp \left( -\frac{\varphi^2}{\frac{2}{V^2} [W_R\cdot G \cdot W_R]}
\right) \,\, .
\end{equation}
where the compact notation $[W_R\cdot G \cdot W_R] = \int dx\, \int
dz\, W_R(x-y)G(x,z)W_R(z-y)$ stands for the {\it filtered} 
or coarse-grained two-point
correlation function associated with the microscopic field. 
Note that $p(\varphi)$ 
in (\ref{zeroth}) is normalized to unity. We will
calculate the asymptotic scaling form of $[W_R\cdot G \cdot W_R]$
below.  Here $G(x,z)$ is the ``variance'' (or ``dispersion'') of the
microscopic field between space--time points $x$ and $z$. More
generally \cite{spde}, $G(x,z)$ is the connected 2--point correlation
function (or propagator), for the field $\phi$, $\langle \phi(x)
\phi(z) \rangle \equiv G(x,z)$.

In the presence of interactions and fluctuations the characteristic
functional $Z[J]$ is modified. It can be obtained as an expansion in
terms of the $n$--point correlation functions for the field
$\phi(\vec{x},t)$, which themselves are corrected due to interactions
and fluctuations. In fact, as is well known \cite{amit}, the
$n$--point functions obey Renormalization Group equations (which
follow \cite{weinberg}, \cite{goldenfeld} from the fact that removal
of divergences in the model introduces an arbitrary scale $\mu$)
describing how $n$--point functions change as the parameters-- 
the couplings $\{g_j\}$ and mass $m$-- in the
model, 
or {\it the scale at which the system is observed}, are
modified. The solution to the RG equations are the so-called
``improved'' $n$-point functions. For the connected $n$--point
correlation function in momentum space, the RG equation in a
mass-independent subtraction procedure is

\begin{equation}\label{callan}
\left[ \mu \frac{\partial}{\partial \mu} + \sum_i \beta_{i}
\frac{\partial}{\partial g_i} 
+ \delta(g)m\frac{\partial}{\partial m}
+\frac{1}{2} n \gamma_{\phi} \right]
G^{(n)}_{Imp}(q; ...)=0\,\,.
\end{equation}
Here, $\gamma_{\phi}$ is the
anomalous dimension of $\phi(x,t)$, given by $\gamma_{\phi}=\mu
(\partial \ln Z_{\phi}/\partial \mu)$ where $Z_{\phi}$ is the wave function
renormalization constant of ${\phi}$, $\delta(g) = -\mu(\partial \ln Z_m/\partial \mu)$
where $Z_m$ is the mass renormalization constant,  
and $q$ represents the momentum and
frequency variables.

A RG 
equation for the improved probability density can be derived 
directly from $Z$ as follows.
Assuming a renormalizable field theory, the relation between the bare
and renormalized generating functional is 
$Z[J_0,\{{g_0}_i\},m_0,\Lambda] = Z[J, \{{g}_i\},m,\mu]$.
The functional written in terms of the bare parameters and bare
source (all having the zero-subscript) and cut-off
$\Lambda$ does not know about the
arbitrary finite scale parameter $\mu$ , so that
$\mu \frac{d Z[J,\{{g}_i\},m,\mu]}{d \mu} = 0$.
The chain-rule immediately implies that 
\begin{equation}\label{chainrule}
\Big(\mu\frac{\partial}{\partial \mu} + \sum_j \beta_j \frac{\partial}{
\partial g_j} + \delta(g_i) m \frac{\partial}{\partial m} + 
\frac{\gamma_{\phi}}{2} 
\,\int d x \, J(x) \, \frac{\delta}{\delta J(x)} \Big)
Z[J,\{{g}_i\},m,\mu] = 0,
\end{equation} 
where the coefficient functions
\begin{eqnarray}\label{beta}
\mu \frac{\partial g_j}{\partial \mu} &=& \beta_j(g),  \\
\label{delta}
\mu \frac{\partial m}{\partial \mu} &=& \delta(g_i)\, m,  \\
\label{source}
\mu \frac{\partial J(x)}{\partial \mu} &=& 
\frac{\gamma_{\phi}}{2} \, J(x),
\end{eqnarray}
describe the scale dependence for the 
couplings $g_j$, the mass $m$ 
(if there is any) and the
source function $J$. 
We emphasize that the renormalization of the source
is equivalent to wavefunction renormalization, an important
fact that is emphasized by Brown \cite{Brown}.  
The source 
function acts as just another bare ``parameter'' of the theory, and it gets
renormalized along with the other couplings.   
This RG equation for $Z[J]$ holds for arbitrary source functions. 
Due to
(\ref{2}) we now substitute    
$J(x) \rightarrow \frac{i\xi}{V} W_R(x)$, 
into (\ref{chainrule}) and 
then by means of the identity (proof: expand Z[J] out in a functional
Taylor series in powers of $J$),
\begin{equation}\label{functional}
\int d x\, J(x)\frac{\delta}{\delta J(x)}Z[J] =
\int d x\, W_R(x)\frac{\delta}{\delta W_R(x)}Z[i\xi W_R/V] =
\xi\frac{\partial}{\partial \xi}Z[i\xi W_R/V],
\end{equation}
we obtain
\begin{equation}\label{chainrule2}
\Big(\mu\frac{\partial}{\partial \mu} + \sum_j \beta_j \frac{\partial}{
\partial g_j} + \delta(g_i) m \frac{\partial}{\partial m} + 
\frac{\gamma_{\phi}}{2} 
\, \xi\frac{\partial }{\partial \xi} \Big)
Z[i\xi W_R/V,\{{g}_i\},m,\mu] = 0.
\end{equation} 
Following (\ref{2}) we Fourier transform this to arrive at the
RG equation for $p_{Imp}$: 
\begin{equation}\label{CSp}
\Big(\mu\frac{\partial}{\partial \mu} + \sum_j \beta_j \frac{\partial}{
\partial g_j} + \delta(g_i) m \frac{\partial}{\partial m} - 
\frac{\gamma_{\phi}}{2} \, \frac{\partial}{\partial \varphi} \varphi 
\Big)
p_{Imp}(\varphi;\{{g}_i\},m,\mu; R) = 0,
\end{equation}
which follows after an integration by parts to eliminate $\xi$-derivatives
in favor
of $\varphi$-derivatives.

\noindent
Just as does equation (\ref{callan}), this equation expresses the
independence of the $physics$ on the choice of scale $\mu \sim
\frac{1}{R}$ at which we defined the values of the coupling constants
$g_i$ of the model. The coefficient functions $\beta_j$, which are
calculable using perturbation theory, describe the scale dependence
for each of the couplings according to (\ref{beta}).
The scale $\mu$ is known as the ``sliding'' scale, and represents the
scale at which the system is sampled. A hurricane looks very different
depending on whether it is seen by a fly trapped inside it or by an
astronaut from outer space; in one case $\mu_{Fly} \sim 1/
\ell_{Fly}$, where $\ell_{Fly}$ is the typical length scale for a fly
and correspondingly for the astronaut, where now $\ell_{Astronaut}$ is
the size of the region of the Earth observed by the astronaut.
$\varphi(\mu)$ satisfies the differential equation $\mu \frac{d
  \varphi(\mu)}{d \mu}=-\frac{1}{2}\gamma_\phi \varphi(\mu)$ 
(compare this to (\ref{source})) and
represents the value of the coarse-grained field at scale $\mu$.
The explicit relation between $\mu$ and $R$ is
discussed in more detail below.

The solutions to equations (\ref{callan}) (for $n=2$) and (\ref{CSp})
are respectively given by 

$$
G_{Imp}(q; ...;g_j(\mu),m(\mu);\mu)=
$$
\begin{equation}\label{impr}
e^{-\int_{\mu_0}^{\mu} \gamma_{\phi}(u)\, d \ln u}\cdot
G_{Imp}(q_0; ...;g_j(\mu_0), m(\mu_0);\mu_0) \, \, ,
\label{10}
\end{equation}

\noindent
and 
\begin{equation}\label{soln}
p_{Imp}(\varphi(\mu),\{{g}_i(\mu)\},m(\mu),\mu;R) =
e^{\frac{1}{2}\int_{\mu_0}^{\mu} \gamma_{\phi}(u) \,d\ln u}\,
p_{Imp}(\varphi(\mu_0),\{{g}_i(\mu_0)\},m(\mu_0),\mu_0 ;R).
\end{equation}

These solutions are to be interpreted as follows: for example, to
obtain the improved \cite{footnote1} 2--point correlation function
$G_{Imp}$, one needs to write down its explicit form at
some scale $\mu_0$ where it is known, and the values of the various
couplings $g_i$ and mass $m$ must be substituted by their ``running'' or
``effective'' value $g_i(\mu)$ and $m(\mu)$ which are the solutions to the RGE
equations (\ref{beta},\ref{delta}).

As one approaches a fixed point of Eq. (\ref{beta}), the couplings on
which the anomalous dimension $\gamma_{\phi}$ depends go to constant
values $g_i^*$, and the anomalous dimension reaches a constant value;
similarly, as a result of equation (\ref{callan}), the two-point
correlation function $G_{Imp}(q; ...;g(\mu);\mu)$ goes
(after Fourier transforming) into a function $f(u)$ of argument,
$u=t/r^z$ (where $z$ is the value at the fixed point of a different
anomalous dimension, the so called dynamical exponent) times $r^{2
  \chi}$, where $\chi$ is related to $\gamma_{\phi}$ at the fixed
point. Here, $r=|\vec x- \vec z|$, and $t=|t_x - t_z|$.  
In the
neighborhood of any RG fixed point,
it is 
easy to derive the scaling form of the Green functions
of the field theory in terms of dynamic critical exponents 
$\chi,z$ as follows.
Under independent rescaling of 
coordinates and the time ${\vec x} = s{\vec x'}$,
$t = s^z t'$ 
(note: this kind of independent space and time scaling is needed when
treating non-relativistic theories such as arise 
for example in diffusion and growth processes) 
the field scales as $\phi(\vec x,t)  =
s^{\chi}\phi({\vec x'},t')$, so it follows that
\begin{eqnarray}\label{Gscale}
G(\vec x,t) &=& \langle \phi({\vec x},t)\phi({\vec 0},0)\rangle \nonumber \\ 
&=& 
s^{2\chi}\, \langle \phi(s^{-1}{\vec x},s^{-z} t)\phi(0,0)\rangle,\nonumber \\
&=& s^{2\chi}\, G(s^{-1}\vec x, s^{-z}t)\nonumber \\
&=& r^{2\chi} \, f\Big(\frac{t}{r^{z}}\Big),
\end{eqnarray}
where the last line follows from 
choosing $s \sim |\vec x|$, $r = |\vec x|$.    
The asymptotic behavior of $f(u)$ is given by \cite{freyandtauber}

\begin{equation}\label{f}
\lim_{u \rightarrow 0} f(u) \rightarrow {\rm const.} \,\, ,
\,\,{\rm and} \qquad
\lim_{u \rightarrow \infty} f(u) \rightarrow u^{2 \chi/z} \,\, .
\end{equation}

To aid our understanding of the relation between the 
coarse-graining scale $R$ and the sliding (momentum) scale $\mu$, 
we can use the above general solution (\ref{soln}) plus simple dimensional
analysis. Let 
$d_{\varphi}$ denote the canonical dimension of the field: 
$[\varphi] = \mu^{d_{\varphi}}$,
then $p(\varphi)$ 
must have dimension $[p] = [\varphi^{-1}] =  
\mu^{-d_{\varphi}}$ expressed in units of 
the sliding scale $\mu$. 
The mass dimension $[m] = \mu$, and assuming dimensionless couplings, then 
$[g_j] = \mu^0 = 1$. Then from (\ref{soln}) it is easy to prove that
\begin{equation}\label{scaling}
p_{Imp}(\varphi(\mu_0),\{{g}_i(\mu_0)\},m(\mu_0),\mu_0 ;R) =
\mu^{-d_{\varphi}}\, 
e^{- \frac{1}{2}\int_{\mu_0}^{\mu}  \,\gamma_{\phi}(u) \, d\ln u}
F \Big(\frac{\varphi(\mu)}{\mu^{d_{\varphi}}}, \frac{m(\mu)}{\mu},
g_j(\mu); \mu R     
\Big),
\end{equation}
where $F$ is a {\it dimensionless} function of the
{\it dimensionless} arguments as wriiten here. Cast in this form, we can investigate
the infrared limit of the probability density by taking $\mu \rightarrow 0$.
The connection to the Wilsonian RG approach and the 
window scale is the following. In the Wilsonian RG, the degrees
of freedom are coarse-grained either in real space or in momentum space, the
latter typically proving to be the more technically convenient {\it choice}.  
This is performed over a finite-width momentum shell corresponding to
$\Lambda/s \leq |\vec k| \leq \Lambda$; this corresponds in fact to
a ``top-hat'' window in momentum space.  
The UV cutoff $\Lambda$ is 
contracted down to $\Lambda/s$, where $s>1$. The infrared limit obtains by taking
$s \rightarrow \infty$. The connection with the field theory sliding
scale is $\mu = \Lambda/s$ \cite{LeBellac}. 
Therefore, when coarse-graining in real space, the contracted 
cutoff corresponds to
an increasing length scale: $R = s/\Lambda$. 
So, the IR limit $\mu \rightarrow 0$ 
corresponds to
$R \rightarrow \infty$. The dimensionless product
$\mu R =1$, and we can now replace $\mu = 1/R$ 
everywhere in 
(\ref{scaling}). In the neighborhood of an IR fixed point, reached by
taking $R \rightarrow \infty$, we 
therefore obtain the asymptotic scaling form:
\begin{eqnarray}\label{scaleform}
\lim_{R \rightarrow \infty} p_{Imp}(\varphi(\mu_0),\{{g}_i(\mu_0)\},&m(\mu_0)&,\mu_0 ;R) 
=
R^{d_{\varphi}+ \frac{1}{2}\gamma_{\phi}(g^*)}\, \nonumber \\ 
&\times&
F \Big( \frac{\varphi(\mu_0)}{\mu_0^{d_{\varphi}}} \big(\frac{\mu}{\mu_0}\big)^{
-d_{\varphi}-\frac{1}{2}\gamma_{\phi}(g^*)}, 
\frac{m(\mu_0)}{\mu_0}\big(\frac{\mu}{\mu_0}\big)^{
\delta(g^*)-1} ,
g_j^* \Big),
\end{eqnarray}
where $\mu_0 = 1/R_0$ is some reference scale and we have used the
solutions of 
(\ref{beta},\ref{delta}) at the 
fixed point $g_j^*$ in arriving at this final form.
At the fixed point, the exponential prefactor in (\ref{scaling}) scales as 
$\big(\mu/\mu_0\big)^{-\frac{1}{2}\gamma_{\phi}(g^*)}$.

Note that (\ref{scaleform}) demonstrates {\it in general} that the large scale
asymptotic form of the 
coarse-grained probability density goes as a non-trivial power of the
window size $R$ times a certain dimensionless function. The asymptotic behavior
is controlled by the canonical dimension $d_{\varphi}$ of the field, its
anomalous dimension at the IR fixed point $\gamma_{\phi}(g^*)$ 
(as well as by $\delta(g^*)$ in a
massive theory). Specification of $p_{Imp}$ at some reference scale 
$R_0 = 1/\mu_0$ yields
the explicit mathematical form of $F$. 
The derivation of the RG equation for $p_{Imp}$ in (\ref{CSp}),
and the large-distance 
scaling behavior of its general solution in (\ref{scaleform})
are the key results of this Letter.

To illustrate the use of the above $general$ results, we apply them to
two simple examples: (i) a free particle in the presence of a heat
bath, and (ii) a system described by the Kardar-Parisi-Zhang (KPZ)
equation with colored noise \cite{mhkz}.

In the case of a free particle in a heat bath there is no self-interaction,
but the statistics of the bath turns the problem into a classical
Brownian motion problem; for the KPZ system there are interactions
among the particles making up the system (described by the KPZ field)
and also interactions with the bath (which could be an external
environment or the effective result of a ``microscopic'' dynamics)
represented by a noise term that drives the time derivative of the KPZ
field. For each of these examples we
will obtain the form of equation (\ref{scaling}) which corresponds to the
scale--dependent form of the probability $p_{Imp}(\varphi; R)$. 
To lowest order, the probability density is gaussian, thus the 
procedure consists in computing the improved form of
$G(\vec{x},\vec{z};t_x,t_z)$ appearing in Eq.  (\ref{zeroth})
and given by Eq. (\ref{10}). 
We indicate briefly how to RG-improve non-gaussian
probabilities below. 

Example (i). 
The equation of motion for a point particle (a particle may be
regarded as a zero-dimensional field) in a medium with viscous
drag and subject to a random force is
\begin{equation}
\label{langevin}
\dot {\vec v}(t) = -\gamma \vec v(t) + \eta(t).
\end{equation}

Here $P$ of Eq. (\ref{3}) is \cite{zinnjustin}

\begin{equation}
P[r] \propto \exp \left\{ -\frac{1}{4} \int_0^t ds \left(\frac{d
      \vec{r}}{ds}\right)^2\right\} \,\, ,
\label{14}
\end{equation}

\noindent
where $\vec{r}(t)$ denotes the coordinate of the particle at time
$t$ and $\vec{v}(t)$ is the particle's velocity. 
The object $G(t,t')$ is

\begin{equation}
G_{ij}(t,t')=\langle r_i(t) r_j(t') \rangle \propto t \,\,
\delta_{ij} \delta(t-t') \,\, .
\label{15}
\end{equation}
                   
There are no corrections due to fluctuations or interactions, and the
probability of Eq. (\ref{zeroth}) is simply given by inserting
Eq. (\ref{15}) into (\ref{zeroth}) (after using a temporal ``top-hat''
window $W_T$)

\begin{equation}
p(r) \propto \frac{1}{T^{d/2}}\cdot \exp \left(
  -\frac{r^2}{2T}\right) \,\, ,
\label{16}
\end{equation}

\noindent
(where $d$ is the number of components of the vector $\vec{r}$)
which of course is the probability distribution that a particle
executing standard Brownian motion be at 
position $r = |\vec r|$ at time $T$.
                                       
Example (ii). 
The KPZ equation (\ref{kpz}) is a non-linear 
Langevin equation for a field. Contrast this to (\ref{langevin})
which is a linear Langevin equation for a 
point-particle.  
In a system described by the KPZ equation

\begin{equation}\label{kpz}+
\frac{\partial \phi}{\partial t}= \nu \nabla^2 \phi
+\frac{1}{2}\lambda 
(\nabla \phi)^2 +\eta(x,t)
\label{17}
\end{equation}

\noindent
with colored noise \cite{footnote2}
$\eta(\vec{x},t)$, there are corrections due to both fluctuations and
interactions. 
Setting $\lambda = 0$ yields the linear
Edwards-Wilkinson (EW) model having a unique IR fixed
point $P1$, for which the IR critical
exponents are known exactly for all space dimensions: 
$(z,\chi) = (2, (2-d)/2)$, in the case of white uncorrelated noise
\cite{Barabasi}. The EW model is the free-field
limit of the KPZ equation and generalizes the concept of random
walk of a classical particle to the level of free fields.
For non-zero $\lambda$ two new fixed points arise 
and their corresponding  critical exponents $\chi$
and $z$ have values differing from the EW fixed point 
exponents.
Briefly, in $d=3$  space dimensions
$P1$ is a saddle point with exponents 
$(z,\chi) = (2,-\frac{1}{2})$,
$P2$ is infrared unstable with $(z,\chi) = (\frac{13}{6},-\frac{1}{6})$,
and $P3$ is infrared stable with $(z,\chi) = (\frac{2}{3},\frac{4}{3})$.
For each of these fixed points one has
to consider two possibilities (cf. Eq.  (\ref{f})
above), depending on whether $|t-t'| \gg |\vec{r}-\vec{r}'|^z$ or
$|t-t'| \ll |\vec{r}-\vec{r}'|^z$ since, as mentioned above, these
limits lead to different asymptotic 
behaviors for the scaling function $f$ in (\ref{f}).  The
corrected form of the PDF that the effective field has $value$
$\bar{\varphi}$ is then obtained by inserting the
``improved'' form of $G$, Eq. (\ref{Gscale}), into
Eq.(\ref{zeroth}); coarse-graining it with the window function $W_R$
and using (\ref{soln}) leads to

$$
p_{Imp}(\bar{\varphi}; R) 
$$
$$
= e^{-\frac{1}{2} \int_{\mu_0}^{\mu} \gamma_{\phi}(u) \, d\ln u}\, 
\sqrt{\frac{V^2}{2\pi[W_R\cdot G_{Imp} \cdot W_R] }}
\cdot \exp \left( -\frac{\bar{\varphi}^2}{\frac{2}{V^2} 
[W_R\cdot G_{Imp} \cdot W_R]}
\right).
$$

To proceed further we need the scaling form of the coarse-grained improved
two-point function. To this end, we take a window function of the
simple form $W_{R,T}(x) = \Theta(R-|\vec x|)\Theta(T-t_x)$ and without loss
of generality, take the window center at the origin. 
(The $scaling$ cannot and does not depend on where the window is located.)
Then we find that
\begin{equation}
\frac{1}{V^2} [W_{R,T} \cdot G_{Imp}\cdot W_{R,T}] \sim
R^{2\chi},
\end{equation}
for $|t| \ll |r|^z$ and 
\begin{equation}
\frac{1}{V^2} [W_{R,T} \cdot G_{Imp}\cdot W_{R,T}] \sim
T^{2\chi/z},
\end{equation}
for $|t| \gg |r|^z$, respectively.

For each of the fixed points 
the asymptotic limits are
              
\begin{equation}
\lim_{|t| \gg |r|^z} p_{Imp}(\bar{\varphi}; R,T)
 \sim T^{-\chi/z+\frac{1}{2}\gamma_{\phi}(g^*)/z}
\exp \left( -\frac{1}{2} \,\, 
\frac{\bar{\varphi}^2}{T^{2\chi/z}}\right)
\label{19}
\end{equation}

\noindent
and

\begin{equation}
\lim_{|t| \ll |r|^z} p_{Imp}(\bar{\varphi}; R, T) 
\sim R^{-\chi + \frac{1}{2}\gamma_{\phi}(g^*)}
\exp \left( -\frac{1}{2} \,\, 
\frac{\bar{\varphi}^2}{R^{2\chi}}\right) \, .
\label{20}
\end{equation}
The roughness exponent and the anomalous dimension are 
related through 
the exponent identity 
$\chi = -d_{\varphi} - \frac{1}{2}\gamma_{\phi}(g^*)$
which follows from using (\ref{impr}) plus dimensional analysis
to arrive at a scaling form for $G_{Imp}$ in complete
analogy to what we worked out above
for $p_{Imp}$ in (\ref{scaling}) and in 
(\ref{scaleform}). Comparing the scaling form 
so obtained with
(\ref{Gscale}) immediately yields this identity.

We see that in the KPZ problem the asymptotic probability
distributions associated with the fixed
points $P2,$ and $P3$ are the ones for $fractal$ Brownian motion
\cite{falconer}, unlike in the free 
{\it field} case, $\lambda = 0$,  or EW model, 
which describes Brownian motion for non-interacting fields.  

In comparing KPZ with EW results, we have fractal Brownian
motion in time, Eq (\ref{19}), and in space Eq. (\ref{20}), because 
the exponent combinations $2 \chi/z$ and $2\chi$ 
appearing within the exponential function are 
different from the EW values at the fixed
points of the RGE P2 and P3. 

Moreover, a nonzero wavefunction
renormalization $\gamma_{\phi}(g^*) \neq 0$ can modify the 
exponents of the power-law prefactors in 
(\ref{19}) and (\ref{20}) away from their naive
canonical values. For EW, the anomalous dimension is
identically zero. 
The reason for the appearance of fractal Brownian
motion in Example (ii) is now obvious: the combined effect of
fluctuations and interactions drives $\chi$ and $z$ away from their
free field theory (EW) values and the probability distribution is
correspondingly shifted from regular Brownian motion to fractal
Brownian motion.

RG-improvement and the asymptotic scaling
of non-gaussian probabilities can be worked out using the 
techniques of this paper. 
For any interacting field theory, one can expand the exact characteristic
functional about the gaussian limit and proceed to derive the associated
RG equation for $p_{Imp}$ 
by substituting
the identity
\begin{equation}
Z[J] = \exp \Big(S_{interaction}[\frac{\delta }{\delta J}] \Big)\,
Z_{Gaussian}[J]|_{J\rightarrow \frac{i\xi}{V} W_R},
\end{equation}
directly into the definition of $p_{Imp}$ in (\ref{2}). 
Here, $S_{interaction}$ represents
the non-Gaussian part of the action. In our examples, we have calculated
the lowest order correction, which is the RG improved Gaussian. Non-Gaussian
terms will appear at higher order in the couplings and these will
take the form of a polynomial in $\varphi$ times a gaussian. The polynomial
will depend on higher $n$-point Green functions 
(e.g., for $n \geq 2$), each of which obeys
Eq(\ref{callan}).

In summary we see that the $combined$ action in space-time evolution
of interactions and fluctuations leads to probability distributions
which at some scale may be gaussian. However, given a particular
dynamics, as the spatial extent of the system or the time window over
which the system is observed, is changed, the two-point correlation
function acquires an anomalous dimension and the effective PDF becomes
the one for fractional Brownian Motion (fBM). This fBM is commonly
associated with complex behavior, such as known to occur in the
financial markets, ecology or in river systems \cite{science99}, and
its character depends on the initial values assigned to the relevant
couplings at some scale. We have developed the general framework and
carried out the above analysis for two complementary problems. For the
free field, as expected, we recover Brownian motion; this is useful as
a very simple test of the validity of the application of our
framework; in the much more complex case of KPZ we find a wealth of
behaviors, with $persistence$ (related to $superdiffusive$ processes)
$antipersistence$ (related to $subdiffusive$ processes) and regular
Brownian Motion \cite{levyflights}. In fact, our analysis explicitly
shows that the nature of the fBM displayed by the field system is
related to the choice of initial conditions for the couplings, since
the values of the roughness and dynamical exponents $\chi$ and $z$
depend on which basin of attraction of the fixed points the initial
couplings are chosen. Thus, the notion of complexity is scale
dependent as well as dependent on the initial conditions; furthermore,
since a given dynamics may have a variety of fixed points, we see
explicitly that a particular system may display various complex
behaviors which reflect the presence of different environments.

Finally, we note that the methods presented here can be extended and
applied to any system which can be described by a field theory, such
as fluids, materials, aggregates, and so forth.

\begin{acknowledgments}
The authors thank M. Visser for discussions. J. P--M thanks
M. Gell--Mann and G. West for many stimulating conversations on the
emergence of power laws in science. 
\end{acknowledgments}


\begin{thebibliography}{99}
\bibitem{footnote1} The term ``improved'' means \cite{gml},
\cite{weinberg} and \cite{goldenfeld}, that the effects of
interactions and fluctuations have been taken into account. This is so
on account of the fact that the coefficients in equations (\ref{callan})
and (\ref{CSp}) are already calculated at each order of perturbation
theory. The objects $p_{Imp}$ and $G^{(2)}_{Imp}$ contain much
more information.
\bibitem{footnote2} The noise is defined to be gaussian, with
  zero mean and a two--point correlation function given by 
$
  \langle \eta (\vec{x},t) \eta(\vec{x}',t')\rangle =2 D(\vec{x},
  \vec{x}';t, t').
$
The local function $D(\vec{x}, \vec{x}';t, t')$ is the double Fourier
transform of $D(\vec{k},\omega)\propto k^{-2 \rho} \omega^{-2
\theta}$. The existence of power law behavior for the correlation
functions of $\phi$ is $not$ due to the choice of noise correlations
given above; it is a $generic$ prediction of the Renormalization Group
due to the fact that the RGEs are quasi--linear
first order partial differential equations.
\end{thebibliography}
\end{document}